\documentclass[onecolumn,useAMS,usenatbib]{mn2e}

\usepackage{graphicx,amsmath,natbib,bm,multirow,color}

\title[3-point intrinsic alignment self-calibration]
{Self-Calibration for 3-point Intrinsic Alignment \emph{Auto}-Correlations in Weak Lensing Surveys}

\author[Troxel \& Ishak]
 {M. A. Troxel\thanks{Electronic address:
    {\tt troxel@utdallas.edu}} and
  M. Ishak\thanks{Electronic address:
    {\tt mishak@utdallas.edu}}$^1$ 
\\$^1$Department of Physics, The University of Texas at Dallas, Richardson, TX 75083, USA
}

\begin{document}

\date{\today}

\pagerange{\pageref{firstpage}--\pageref{lastpage}} \pubyear{0000}

\maketitle

\label{firstpage}

\begin{abstract}
The weak lensing signal (cosmic shear) has been shown to be strongly contaminated by the various types of galaxy intrinsic alignment (IA) correlations, which poses a barrier to precision weak lensing measurements. The redshift dependence of the IA signal has been used at the 2-point level to reduce this contamination by only measuring cross-correlations between large redshift bins, which significantly reduces the galaxy intrinsic ellipticity - intrinsic ellipticity (II) correlation. A self-calibration technique based on the redshift dependences of the IA correlations has also been proposed as a means to remove the 2-point IA contamination from the lensing signal. We explore here the redshift dependences of the IA and lensing bispectra in order to propose a self-calibration of the IA auto-correlations at the 3-point level (i.e. GGI, GII, and III), which can be well understood without the assumption of any particular IA model. We find that future weak lensing surveys will be able to measure the distinctive IA redshift dependence over ranges of $|\Delta z^P|\le 0.2$. Using conservative estimates of photo-z accuracy, we describe the 3-point self-calibration technique for the total IA signal, which can be accomplished through lensing tomography of photo-z bin size $\sim 0.01$. We find that the 3-point self-calibration can function at the accuracy of the 2-point technique with modest constraints in redshift separation. This allows the 3-point IA auto-correlation self-calibration technique proposed here to significantly reduce the contamination of the IA contamination to the weak lensing bispectrum.
\end{abstract}


\section{Introduction}\label{intro}

Originally just a prediction of general relativity, gravitational lensing has recently emerged as an independent means to make precision astrophysical and cosmological measurements, which are blind to the exact nature of the lens mass. This is particularly true for weak lensing due to large scale structure (cosmic shear), which is able to map structure composed of both visible matter as well as dark matter, which is primarily detectable through its gravitational signal only. The importance of this new probe has spurred the development of a new generation of ground- and space-based surveys suited for precision weak lensing measurements. These ongoing, future, and proposed surveys (e.g. CFHTLS ({\slshape http://www.cfht.hawaii.edu/Science/CFHLS/}), DES ({\slshape http://www.darkenergysurvey.org/}), EUCLID ({\slshape http://sci.esa.int/euclid/}), HSC ({\slshape http://www.naoj.org/Projects/HSC/}), HST ({\slshape http://www.stsci.edu/hst/}), JWST ({\slshape http://www.jwst.nasa.gov/}), LSST ({\slshape http://www.lsst.org/lsst/}), Pan-STARRS

\noindent({\slshape http://pan-starrs.ifa.hawaii.edu/}), and WFIRST ({\slshape http://wfirst.gsfc.nasa.gov/})) promise to provide greatly improved measurements of cosmic shear using the shapes of up to billions of galaxies. These cosmic measurements allow us to not only characterize the equation of state of dark energy, but when combined with other probes can improve constraints on the equation of state of dark energy and the matter fluctuation amplitude parameter by factors of 2 to 4 (see for example \cite{1p,1c,1f,1a,1n,1l,1e,1g,1o,1b,1h,1k,ishak,1j,1u,7i,1d,1i,1m} and references therein.) Weak lensing has also been shown to be very useful to test the nature of gravity at cosmological distance scales (see for example the partial list \cite{2l,2c,2h,2p,2g,2j,2o,2a,2d,2k,2i,2m,2q,2b,2e,2n,2f}.)

The 3-point cosmic shear correlation and the shear bispectrum have been shown to break degeneracies in the cosmological parameters in addition to the constraints obtained from the 2-point cosmic shear correlation and the corresponding shear power spectrum that the power spectrum alone does not \citep{3a,3b}. For example, the results of \cite{tj} showed that a deep lensing survey should be able to improve the constraints on the dark energy parameters and the matter fluctuation amplitude by a further factor of 2-3 using the bispectrum. Most recently, \cite{5} derived parameter constraints by measuring the third order moment of the aperture mass measure using weak lensing data from the HST COSMOS survey. They found independent results consistent with WMAP7 best-fit cosmology and an improved constraint when combined with the 2-point correlation. In addition to improved parameter constraints, by definition the bispectrum also allows us to explore information about non-Gaussianity in the universe that is inaccessible at the 2-point level, providing constraints on the degree of non-Gaussianity (see for example \cite{matarrese,verde,tj,28,huterer,munshi} and references therein.) 

\begin{figure}
\center
\includegraphics[scale=0.6]{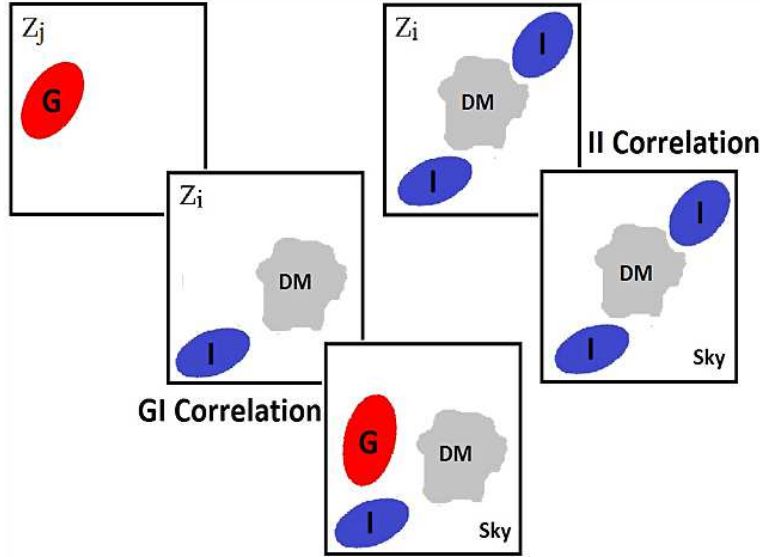}%
\caption{The 2-point intrinsic alignment correlations. Galaxies which are intrinsically aligned are coloured in blue and labeled I, while galaxies which are lensed are coloured in red and labeled G. The lower right panels represent the view of the system on the sky, while each panel preceding it is at some distinct redshift where $z_i<z_j$. If the two galaxies are spatially close, at nearly the same redshift and angular position on the sky, they can be aligned by the tidal force field of the same nearby matter structure (labeled DM in the figure). This is shown as the II correlation. If instead a matter structure causes both the alignment of a nearby galaxy and contributes to the lensing signal of a background galaxy, this produces an anti-correlation with negative sign between the cosmic shear and intrinsic ellipticity, since the tidal force and gravitational lensing tend to align the galaxy shapes in orthogonal directions, and is shown as the GI correlation.}\label{fig:2ptia}
\end{figure}

As we describe below, we extend in this paper the 2-point self calibration technique proposed by \cite{23} to the 3-point intrinsic alignment auto-correlation bispectra between galaxies in a single redshift bin. This technique is different from the cross-correlation techniques proposed in \cite{22} and \cite{troxel}, instead using differences in the redshift dependencies of the intrinsic alignment and lensing bispectra to self-calibrate the intrinsic alignment signal. These 2-and 3-point intrinsic alignment correlations constitute a contaminant to the lensing signal and must be isolated and removed to avoid biasing the cosmological information contained within the cosmic shear power spectra and bispectra.

Cosmic shear measurements are in fact limited in precision by several systematic effects which must be accounted for in order to make full use of the potential of future weak lensing surveys (see for example \cite{6d,6f,6a,6c,6e,6b,7,6j,6h,6k,6m,6g,6i,6l} and references therein), and one of the serious systematic effects of weak lensing is this correlated intrinsic alignment of galaxy ellipticities, which act as a nuisance factor (see for example \cite{6d,6c,8c,7,8g,6j,8e,8f,18b,hirata04,17,13,12,bk,10,semb,14a,14b,9b,15,19,8a,8h,troxel2} and references therein). For example, \cite{bk} and \cite{9b} showed that if intrinsic alignment is ignored, the determination of the dark energy equation of state is biased by as much as $50\%$. \cite{10} found that the matter power spectrum amplitude can be affected by intrinsic alignment by up to $30\%$, showing the importance of developing methods to isolate the intrinsic alignment and remove it from the cosmic shear signal.

\begin{figure}
\center
\includegraphics[scale=0.7]{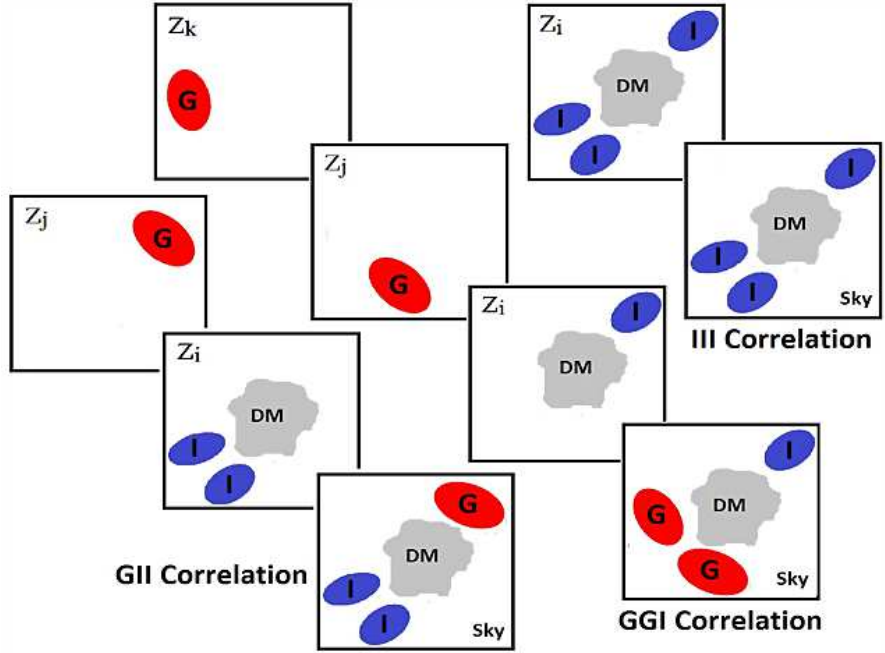}%
\caption{The 3-point intrinsic alignment correlations. Galaxies which are intrinsically aligned are colored in blue and labeled I, while galaxies which are lensed are colored in red and labeled G. The lower right panels represent the view of the system on the sky, while each panel preceding it is at some distinct redshift where $z_i<z_j<z_k$. The III correlation is between the intrinsic ellipticities of three spatially close galaxies which are intrinsically aligned by a nearby matter structure (labeled DM in the figure). If instead two spatially close galaxies are intrinsically aligned by a nearby matter structure which contributes to the lensing of a third galaxy in the background, we label this the GII correlation. Finally, the GGI correlation is where two galaxies are lensed by a structure which intrinsically aligns a third galaxy in the foreground. Unlike the 2-point correlations, the sign of the GGI and GII correlations can depend both on triangle shape and scale.}\label{fig:3ptia}
\end{figure}

There are two 2-point intrinsic alignment correlations. The first is a correlation between the intrinsic ellipticity of two galaxies, known as the II correlation. If the two galaxies are spatially close, they can be aligned by the tidal force field of the same nearby matter structure. The second intrinsic alignment correlation, known as the GI correlation, was identified by \cite{hirata04} and is due to a matter structure both causing the alignment of a nearby galaxy and contributing to the lensing signal of a background galaxy. This produces an anti-correlation between the cosmic shear and intrinsic ellipticity, since the tidal force and gravitational lensing tend to align the galaxy shapes in orthogonal directions. We represent these correlations diagrammatically in Fig. \ref{fig:2ptia}. The GI correlation has been measured in various subsets of the SDSS spectroscopic and imaging samples by various groups. A detection of the large-scale GI correlation in the SDSS was reported by \cite{12}, and then \cite{10} found an even stronger GI correlation for Luminous Red Galaxies (LRGs). It was shown in these papers that this contamination can affect the lensing measurement and cosmology up to the $10\%$ level and up to $30\%$ in some cases for the matter fluctuation amplitude. This finding was confirmed by numerical simulations, where a level of contamination of $10\%$ was found \citep{13}. Further measurements of the GI correlation were made in the SDSS dataset by \cite{14a} and \cite{14b}. Most recently, \cite{15} measured strong 2-point intrinsic alignment correlations in various SDSS and MegaZ-LRG samples.

In a similar way, when we consider three galaxies and the related 3-point correlation, the cosmic shear signal (GGG bispectrum) also suffers from contamination by the 3-point intrinsic alignment correlations. The first is the III correlation between intrinsic ellipticities of three spatially close galaxies which are intrinsically aligned by a nearby matter structure. The second is the GII correlation, where two spatially close galaxies are intrinsically aligned by a nearby matter structure which contributes to the lensing of a third galaxy in the background. Finally, there is the GGI correlation, where two galaxies are lensed by a structure which intrinsically aligns a third galaxy in the foreground. Unlike the 2-point correlations, the sign of the GGI and GII correlations depend both on triangle shape and scale. The 3-point intrinsic alignment correlations are represented diagrammatically in Fig. \ref{fig:3ptia}. \cite{semb} showed that lensing bispectrum measurements are typically more strongly contaminated by intrinsic alignment compared to the lensing spectrum measurements, and that the contamination from the 3-point intrinsic alignment correlation can be as large as $15-20\%$ compared to the GGG lensing signal. The 3-point intrinsic alignment measurements are not only useful for constraining their contamination to 3-point lensing measurements, but are also useful for constraining models of intrinsic alignments and therefore constraining the contamination to all lensing measurements (including 2-point correlations) which will dominate the science cases of upcoming surveys.

While the II, III, and GII intrinsic alignment correlations can be greatly reduced with photo-z's by using cross-spectra of galaxies in two different redshift bins (see for example \cite{6k}) so that the galaxies are separated by large enough distances to assure that the tidal effect is weak, this does not work for the GI and GGI correlations which remain strong between galaxies at different redshifts and large separations. The GI correlation and methods for its removal have been the topic of several recent scientific publications and we review these briefly. Initially, some first suggestions were discussed by \cite{hirata04}. \cite{17} extended the approach of template fitting by \cite{18b} to include a treatment of the GI correlation. \cite{bk} and \cite{9b} investigated the effects of the GI correlation on cosmological parameter constraints by assuming a model of the GI intrinsic alignment that is binned in redshift and angular frequency with some free parameters that are marginalized over. \cite{19} performed a cosmological constraint analysis where modeling of intrinsic alignment was included, showing a significant effect on the amplitude of matter fluctuations. Using a geometrical approach, \cite{20a}, \cite{20b}, and \cite{20c} proposed a nulling technique to remove the GI intrinsic alignment contribution by exploiting the redshift dependence of the correlations, but it was found that the technique throws out some of the valuable lensing signal. Most recently, the nulling technique has been applied at the 3-point level for the GGI correlation, but again with similar signal loss to that at the 2-point level \citep{21}. Finally, \cite{22} proposed a technique to self-calibrate the GI intrinsic alignment signal by using additional galaxy density (cross-)correlations which are already present in weak lensing survey measurements. This approach was successfully extended in recent work by \cite{troxel} to the GGI cross-correlation bispectrum. \cite{9b} applied an approach like the self-calibration, using correlations between lensing, intrinsic alignment, number density, and magnification effects to constrain cosmological parameters. They found that the extra information from the additional correlations can make up for the additional free parameters in the intrinsic alignment so that the contamination can be removed without loss of constraining power.

Most recently, \cite{23} showed that redshift dependencies of the intrinsic alignment spectra can allow further improvements to the calculation of the intrinsic alignment contamination. \cite{23} demonstrates the strong redshift separation dependence of the 2-point GI and II intrinsic alignment signals by considering the lensing and intrinsic alignment spectra due to galaxies in a single redshift bin that are at fixed photo-z separations $\Delta z^P$ relative to some mean redshift. This results in a relative change in magnitude of the GI and II spectra of 50-60\%, but only a few percent for GG at separation $\Delta z^P=0.2$. This corresponds to a decrease in 10\% of the total ellipticity power spectrum, which is identifiable in proposed surveys. Parameterizing the intrinsic alignment spectra as proportional to various galaxy density (cross-)spectra, this allows us to calculate and remove the intrinsic alignment component of the lensing signal. We propose in this work a method to extend this self-calibration technique to the 3-point statistics, using instead the redshift dependencies of the GGI, GII, and III bispectra, in order to calculate and remove the intrinsic alignment contamination to the lensing bispectrum.

We organize the paper as follows. In Sec. \ref{dzpb}, we briefly discuss the necessary survey parameters and lensing formalism for the bispectrum. In Sec. \ref{dzp}, we discuss several methods to vary the redshift separation of galaxies in the bispectrum and compare the resulting redshift dependencies. Section \ref{dep} describes the analytical motivation and framework for the self-calibration using the redshift dependencies of the GGG, GGI, GII, and III bispectra. We explore the performance of the self-calibration in Sec. \ref{perf}, and discuss possible additional sources of error in the analysis in Sec. \ref{other}. Finally, we summarize the self-calibration and its applicability to future weak lensing surveys in Sec. \ref{cal} and discuss its impact in Sec. \ref{conc}. 

\section{Redshift separation dependence in the bispectrum}\label{dzpb}

The galaxy intrinsic alignment and cosmic shear bispectra have very distinct dependences on the redshift separation and orientation of the three galaxies, which we will explore and quantify in order to identify and measure the contribution of the intrinsic alignment to the measured ellipticity bispectrum. We work with the bispectrum $B^{\alpha\beta\gamma}(\ell;z^P_1,z^P_2,z^P_3)$, where $\alpha,\beta,\gamma$ $\in$ $G,I,g$ measures $\alpha$ at photometric redshift $z^P_1$, $\beta$ at $z^P_2$, and $\gamma$ at $z^P_3$. We have denoted the lensing convergence as \emph{G}, the galaxy intrinsic ellipticity as \emph{I}, and the galaxy density as \emph{g}. Since we are only interested in exploring the redshift separation and orientation dependence of the bispectra, we will work at a fixed multipole $\ell$ and mean redshift $\bar{z}^P=(z^P_1+z^P_2+z^P_3)/3$. In order to best compare to work on the power spectrum by \cite{23}, we will use equilateral triangles with $\ell=1000$ and $\bar{z}^p=1.0$, unless otherwise stated, for which values the bispectrum will be measured to high confidence in planned weak lensing surveys. Unlike for the power spectrum, there are several possible choices in exploring the dependence on redshift separation, which we will discuss and compare in section \ref{dzp}.

We assume in our calculations a standard, flat $\Lambda$CDM universe. In the Born approximation, the convergence $\kappa$ of a source galaxy at comoving distance $\chi_G$ and direction $\hat{\theta}$ is then related to the matter density $\delta$ through the lensing kernel $W_L(z',z)$ by
\begin{equation}
\kappa(\hat{\theta})=\int_0^{\chi_G}\delta(\chi_L,\hat{\theta})W_L(\chi_L,\chi_G)d\chi_L.
\end{equation}
The 3D matter bispectrum is then defined from the convergence as
\begin{equation}
\langle \tilde{\kappa}(\bm{\ell_1})\tilde{\kappa}(\bm{\ell_2})\tilde{\kappa}(\bm{\ell_3})\rangle=(2\pi)^2\delta^{D}(\bm{\ell_1}+\bm{\ell_2}+\bm{\ell_3})B(\ell_1,\ell_2,\ell_3),\label{eq:corr}
\end{equation}
where $\langle\cdots\rangle$ denotes the ensemble average and $\delta^{D}(\bm{\ell})$ is the Dirac delta function. For the bispectrum, $\delta^{D}(\bm{\ell}_1+\bm{\ell}_2+\bm{\ell}_3)$ enforces the condition that the three vectors form a triangle in Fourier space. Under the Limber approximation, we can express the 2D angular auto-correlation bispectrum as
\begin{equation}
B^{\alpha\beta\gamma}(\ell;z^P_1,z^P_2,z^P_3)=\int_0^{\chi}\frac{W^{\alpha\beta\gamma}(\chi';\chi_1,\chi_2,\chi_3)}{\chi'^4}B_{\alpha\beta\gamma}(k;\chi')d\chi',\label{eq:bs}
\end{equation}
where for example when $\alpha=\beta=\gamma=G$, $B_{GGG}(k;\chi')$ is the 3D matter bispectrum shown in Eq. \ref{eq:corr}. However, generally $\alpha,\beta,\gamma\in G,I,g$, where the additional intrinsic alignment (I) and galaxy (g) bispectra are calculated as described further in Sec. \ref{dzp}. The redshift is simply related to $\chi$ through the Hubble parameter, $H(z)$. We can then write the weighting function in terms of redshift as 
\begin{eqnarray}
W^{\alpha\beta\gamma}(z(\chi);z^P_1(\chi_1),z^P_2(\chi_2),z^P_3(\chi_3))&\equiv&W^{\alpha}(z,z^P_1)W^{\beta}(z,z^P_2)W^{\gamma}(z,z^P_3),\label{eq:w}\\
W^G(z,z^P)&=&\int_0^{\infty}W_L(z',z)p(z'|z^P)dz',\\
W^I(z,z^P)&=&W^g(z,z^P)=p(z|z^P),
\end{eqnarray}
where $p(z|z^P)$ is the photo-z probability redshift distribution (PDF). In order to quantitatively examine the redshift separation dependence of the bispectra, we assume a specific form of the PDF modeled after \cite{mabern} with which to describe the photo-z uncertainty in a future weak lensing survey
\begin{eqnarray}
p(z|z^P)=\frac{1-p_{\textrm{cat}}}{\sqrt{2\pi}\sigma(z^P)}\exp\left[\frac{(z-z^P)^2}{2\sigma^2(z^P)}\right]\label{eq:pdf}
+\frac{p_{\textrm{cat}}}{\sqrt{2\pi}\sigma(z^P)}\exp\left[\frac{(z-f_{\textrm{bias}}z^P)^2}{2\sigma^2(z^P)}\right].
\end{eqnarray}
The fraction of outlier galaxies is given by $p_{\textrm{cat}}$, with a true redshift which is biased by a factor $f_{\textrm{bias}}$. We adopt similar values to Zhang, with $\sigma(z^P)=0.05(1+z^P)$, $p_{\textrm{cat}}=0.02$, and $f_{\textrm{bias}}=0.5$. These parameters are chosen to represent the expected photo-z accuracy of a stage IV weak lensing survey at $z^P\approx 1$. We maintain the very conservative fraction of catastrophic outliers used by Zhang, which is a factor of 20 greater than the required fraction in a stage IV dark energy survey, where $p_{\textrm{cat}}<0.1\%$ \citep{bh,hearin}. Typically we assume that $B^{GGI}(\ell;z_1,z_2,z_3)$ is zero for $z_1,z_2<z_3$ and that $B^{GII}(\ell;z_1,z_2,z_3)$ is zero for $z_1<z_2,z_3$ or $z_2\ne z_3$, due to lensing geometry and the redshift separation and orientation dependence of the intrinsic alignment signal discussed in detail in Sec. \ref{dzp} \citep{troxel}. However, due to sometimes large photo-z error, this is not always the case, so when referencing the GGI and GII correlations, we will instead work with the sums $B^{GGI}+B^{GIG}+B^{IGG}$ and $B^{GII}+B^{IGI}+B^{IIG}$, respectively. 

\subsection{Evaluating the redshift separation dependence}\label{dzp}

We calculate the required lensing, intrinsic alignment, and galaxy bispectra using the relations and methods developed in \cite{troxel} and \cite{troxel2}. We assume a deterministic galaxy bias for the galaxy bispectra, while the intrinsic alignment signal is calculated using the model of \cite{sb09} (SB09), which is based on the halo model prescription. We have used the fiducial parameters of their fitting formulae as listed in their Tables I \& II, with $C_1$ estimated by comparison to Fig. 2 of \cite{hirata04}. By design, this model reduces to the linear alignment model of \cite{hirata04} at large scale, but aims for a more motivated model of intrinsic alignment at small scales. For comparison, we also use the toy model adopted by \cite{23}, where the intrinsic alignment spectrum has a simple bias $b^I(k,z)\propto (1+\Delta^2_m(k,z))^a$ ($a\in [0,1/2]$), where $\Delta^2_m(k,z)$ is the three-dimensional matter power spectrum. This allows us to evaluate the behavior of the redshift separation and orientation dependence of the bispectrum for a wide range of intrinsic alignment dependence on both scale and redshift.

We relate the flat-sky bispectrum to the all-sky bispectrum through the Wigner-3j symbol, where
\begin{equation}
B^{\alpha\beta\gamma}_{ijk,\ell_1\ell_2\ell_3}\approx\begin{pmatrix}\ell_1&\ell_2&\ell_3\\0&0&0\end{pmatrix}\sqrt{\frac{(2\ell_1+1)(2\ell_2+1)(2\ell_3+1)}{4\pi}}B^{\alpha\beta\gamma}_{ijk}(\ell_1,\ell_2,\ell_3).
\end{equation}
We calculate the Wigner-3j symbol following the approximation given in Appendix A of \cite{tj}. We then compute the three-dimensional bispectrum due to nonlinear gravitational clustering, $B_{\delta}(k_1,k_2,k_3;\chi)$, following the fitting formula of \cite{sc} with coefficients $F^{\textrm{eff}}_2(k_1,k_2)$ described in Section 2.4.3 of \cite{tj},
\begin{equation}
B_{\delta}(k_1,k_2,k_3;\chi)=2F^{\textrm{eff}}_2(k_1,k_2)P_{\delta}(k_1;\chi)P_{\delta}(k_2;\chi)+2\textrm{ perm.}
\end{equation}
In order to approximate the three-dimensional intrinsic alignment bispectra, we make a direct expansion of this method, using the intrinsic alignment power spectra instead of the nonlinear matter power spectrum, where
\begin{eqnarray}
B_{\delta\delta\tilde{\gamma}^I}(k_1,k_2,k_3;\chi)&=&2F^{\textrm{eff}}_2(k_1,k_2)P_{\delta\tilde{\gamma}^I}(k_1;\chi)P_{\delta}(k_2;\chi)+2F^{\textrm{eff}}_2(k_2,k_3)P_{\delta}(k_2;\chi)P_{\delta\tilde{\gamma}^I}(k_3;\chi)\\\nonumber
&&+2F^{\textrm{eff}}_2(k_3,k_1)P_{\delta\tilde{\gamma}^I}(k_3;\chi)P_{\delta\tilde{\gamma}^I}(k_1;\chi)\\
B_{\delta\tilde{\gamma}^I\tilde{\gamma}^I}(k_1,k_2,k_3;\chi)&=&2F^{\textrm{eff}}_2(k_1,k_2)P_{\tilde{\gamma}^I}(k_1;\chi)P_{\delta\tilde{\gamma}^I}(k_2;\chi)+2F^{\textrm{eff}}_2(k_2,k_3)P_{\delta\tilde{\gamma}^I}(k_2;\chi)P_{\delta\tilde{\gamma}^I}(k_3;\chi)\\\nonumber
&&+2F^{\textrm{eff}}_2(k_3,k_1)P_{\tilde{\gamma}^I}(k_3;\chi)P_{\delta\tilde{\gamma}^I}(k_1;\chi)\\
B_{\tilde{\gamma}^I}(k_1,k_2,k_3;\chi)&=&2F^{\textrm{eff}}_2(k_1,k_2)P_{\tilde{\gamma}^I}(k_1;\chi)P_{\tilde{\gamma}^I}(k_2;\chi)+2\textrm{ perm.}
\end{eqnarray}
We find that this treatment gives reasonable results for the intrinsic alignment bispectra.

\begin{table}
\center
\begin{tabular}{ l c c c c c c }
\hline
 & \multicolumn{2}{c}{Method 1} & \multicolumn{2}{c}{Method 2} & \multicolumn{2}{c}{Method 3} \\
 $\Delta z^P$ & 0.1 & 0.2 & 0.1 & 0.2 & 0.1 & 0.2 \\
\hline
GGG & 0.5\% & 1.5\% & 0.5\% & 1.5\% & 0.5\% & 2\% \\
GGI & 12\% & 48\% & 12\% & 37\% & 20\% & 89\%  \\
GII & 1\% & 6\% & 5\% & 32\% & 8\% & 42\% \\
III & 23\% & 64\% & 28\% & 72\% & 30\% & 77\%  \\
\hline
\end{tabular}
\caption{The $\Delta z^P$ dependence of $B^{GGG}$ (GGG), $B^{GGI}+B^{GIG}+B^{IGG}$ (GGI), $B^{GII}+B^{IGI}+B^{IIG}$ (GII), and $B^{III}$ (III) for the three methods of varying redshift separation described in Sec. \ref{dzp}. The values listed are the ratio $|B(\Delta z^P)/B(\Delta z^P=0)|$, showing the relative change in magnitude between $\Delta z^P=0.0$ and $\Delta z^P=0.1$ and between $\Delta z^P=0.0$ and $\Delta z^P=0.2$. Method 1 shows a similar $\Delta z^P$ dependency to the power spectrum, while Method 3 clearly displays the most distinct $\Delta z^P$ dependency.}
\label{table}
\end{table}

We explore several methods of varying the redshift separation and orientation in the bispectrum. We can define the bispectrum triangle configuration by the redshift of its vertices ($z^P_1$, $z^P_2$, and $z^P_3$) and the $\ell$-modes corresponding to the angular separation of its sides. From these values, we can derive the mean redshift ($\bar{z}^P$) and the side lengths measured in redshift of each triangle ($\Delta z^P_1$, $\Delta z^P_2$, and $\Delta z^P_3$), where the side $\Delta z^P_n$ is defined opposite the vertex $n$ with photo-z $z^P_n$. In all cases, we keep $\bar{z}^P$ constant. The first (Method 1) is most similar to the method employed by Zhang for the power spectrum. We set $z^P_2=\bar{z}^P$ constant, and instead vary $z^P_1$ and $z^P_3$ such that $\Delta z^P_1=\Delta z^P_3=\Delta z^P_2/2$. The quantity $\Delta z^P_2$ is then the separation that we vary, which completely defines the redshift configuration of the triangle in this method.

In Methods 2 and 3, we set $z^P_1=z^P_2$ such that $\Delta z^P_3=0$. We then vary $\Delta z^P_1=\Delta z^P_2$, which completely defines the triangle configuration. In the first case, Method 2, we vary the separation $\Delta z^P_1=\Delta z^P_2$ with $z^P_3$ increasing in redshift. In the second case, Method 3, we instead vary the separation $\Delta z^P_1=\Delta z^P_2$ with $z^P_3$ decreasing in redshift. We can also explore more complicated methods where $\Delta z^P_3$ is constant and non-zero. We then vary $\Delta z^P_2=|\Delta z^P_3-\Delta z^P_1|$. These more complicated cases do still show a large change in the intrinsic alignment bispectra with varying $\Delta z^P_2$, but are less useful for a self-calibration of the intrinsic alignment signal as there is not a simple relationship between the intrinsic alignment and galaxy bispectra as we discuss in Sec. \ref{cal}.

We summarize the $\Delta z^P$ dependence of Methods 1-3 in Table 1, showing the percent change in the magnitude of the bispectra from $\Delta z^P=0.0$ to $\Delta z^P=0.1$ and from $\Delta z^P=0.0$ to $\Delta z^P=0.2$. With Method 1 we find a resulting $\Delta z^P$ dependence in the GGI and III bispectra which is similar to that found by Zhang for the GI and II power spectra, though with almost no detectable dependence in the GII bispectrum. This similarity and the reduced dependence in Method 1 when compared to Methods 2 and 3, which are described below, can be understood from the way we have varied the redshifts in this case, keeping one vertex at a fixed redshift $\bar{z}^P$, which is very similar to the 2-point case where only two points are varying in redshift. Both Methods 2 and 3 show a noticeable increase in $\Delta z^P$ dependence, though the GGI bispectrum in Method 2 is actually less dependent than in Method 1. All three demonstrate a nearly identical percent change over a separation of $\Delta z^P=0.2$ in the GGG bispectrum, at the 1-2\% level.

Method 3 clearly has the strongest $\Delta z^P$ dependence. We find this is significantly stronger than for the 2-point case and both the previous Methods, and so we will focus our discussion on Method 3 as the best candidate for a self-calibration of the intrinsic alignment signal. As described in Table 1, we find a relative change of 8, 20, and 30\% in the GII, GGI, and III bispectra, respectively, between $\Delta z^P=0.0$ and $\Delta z^P=0.1$. However, this increases greatly such that between $\Delta z^P=0.0$ and $\Delta z^P=0.2$, we find a very large relative change of 42, 77, and 89\% in the GII, III, and GGI bispectra, respectively. The change in the GGG bispectrum is only 2\% between $\Delta z^P=0.0$ and $\Delta z^P=0.2$, by comparison. This very strong $\Delta z^p$ dependence will allow us to develop a means to self-calibration of the intrinsic alignment signal, which we describe in the following sections.

\begin{figure}
\center
\includegraphics[angle=270,scale=0.65]{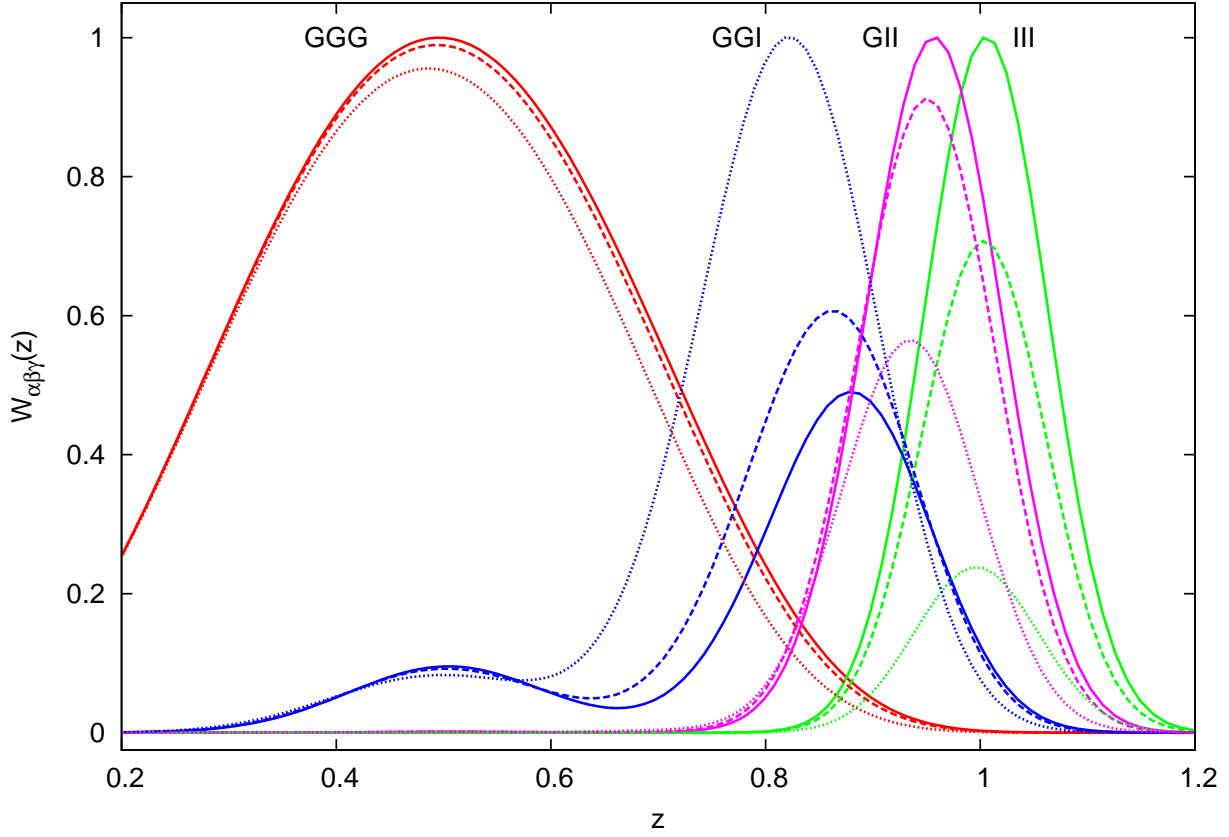}%
\caption{The weighting functions $W^{\alpha\beta\gamma}$ evaluated at $\ell=1000$ and $\bar{z}^P=1.0$. The lines labeled GGG correspond to $W^{GGG}$, GGI to the sum $W^{GGI}+W^{GIG}+W^{IGG}$, GII to the sum $W^{GII}+W^{IGI}+W^{IIG}$, and III to $W^{III}$. The solid lines are evaluated with $\Delta z^P=0.0$, the dashed lines with $\Delta z^P=0.1$ and the dotted lines with $\Delta z^P=0.2$. There is a non-negligible contribution to the sum $W^{GGI}+W^{GIG}+W^{IGG}$ from the fraction $p_{\textrm{cat}}$ of outlier galaxies at $z=0.5$.}\label{fig:wabc}
\end{figure}

\subsection{Describing the redshift separation dependence}\label{dep}

The $\Delta z^p$ dependence in the bispectrum as shown in Eq. \ref{eq:bs} should primarily be determined by the weighting function $W^{\alpha\beta\gamma}$, which has an explicit dependence on $z^P_1$, $z^P_2$, and $z^P_3$. We find this to be particularly true of the ratio $B(\Delta z^P)/B(\Delta z^P=0)$. This observation forms the basis for understanding the $\Delta z^p$ dependence in $B^{GGG}$ (Sec. \ref{ggg}), $B^{GGI}$ (Sec. \ref{ggi}), $B^{GII}$ (Sec. \ref{gii}), and $B^{III}$ (Sec. \ref{iii}). We show in Fig. \ref{fig:wabc} the behavior of $W^{\alpha\beta\gamma}$ as a function of true redshift for various values of $\Delta z^p$.

\subsubsection{Redshift separation dependence in $B^{GGG}$}\label{ggg}

We see from Fig. \ref{fig:wabc} that $W^{GGG}$, though strongly dependent on $\bar{z}^P$, has very little dependence on $\Delta z^p$, which explains the very small change in $B^{GGG}$ as a function of $\Delta z^p$. In the 2-point case, Zhang describes the shear power spectrum with a Taylor expansion about $\Delta z^P=0$ up to second order, where the first derivative of the lensing spectrum with respect to $\Delta z^P$ is shown to be zero. In the case of Method 3, it is not true that the first derivative is zero in the same way due to asymmetry. However, we can apply the same argument to our Method 1, which is symmetric, and express $B^{GGG}$ as

\begin{eqnarray}
\frac{B^{GGG}(\Delta z^P)}{B^{GGG}(\Delta z^P=0)}&\approx& 1-f_{GGG}(\Delta z^P)^2,\label{eq:taylor}\\
f_{GGG}&\equiv & \frac{\partial^2B^{GGG}(\Delta z^P)/\partial(\Delta z^P)^2|_0}{B^{GGG}(\Delta z^P=0)}.\nonumber
\end{eqnarray}

\begin{figure}
\center
\includegraphics[angle=270,scale=0.65]{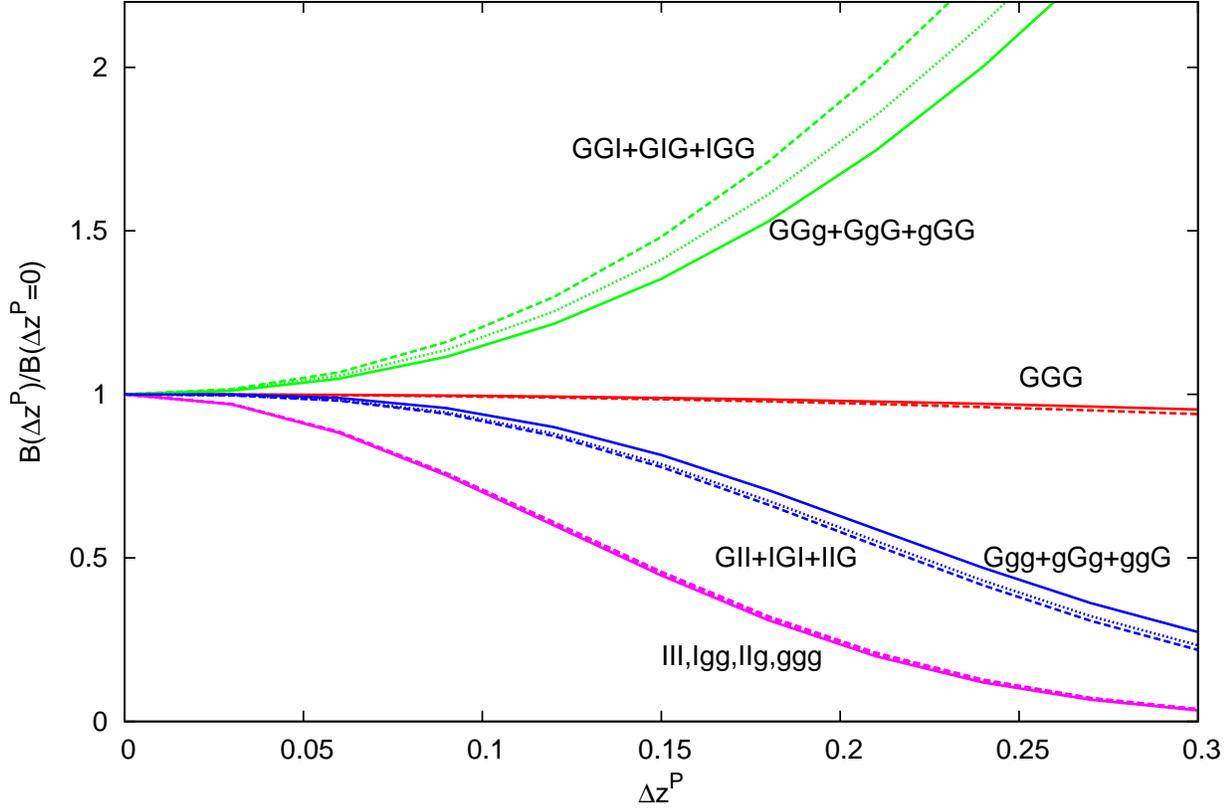}%
\caption{The ratio $B^{\alpha\beta\gamma}(\Delta z^P)/B^{\alpha\beta\gamma}(\Delta z^P=0)$ evaluated for the labeled $\alpha\beta\gamma$ at $\ell=1000$ and $\bar{z}^P=1.0$. This demonstrates the accuracy of the relations in Eqs. \ref{eq:taylor} \& \ref{eq:gggiii}-\ref{eq:ggirel}. The lines labeled GGG which overlap are the left (solid) and right (dashed) sides of Eq. \ref{eq:taylor}. The right side is evaluated as described in Sec. \ref{ggg}. The intrinsic alignment signal is evaluated using both the SB09 model and the toy modal with $a=1/2$. For GGI+GIG+IGG and GII+IGI+IIG, the dashed lines use the SB09 model while the dotted lines use the toy model. The lines labeled III, IIg, Igg, and ggg overlap due to the high accuracy of Eq. \ref{eq:gggiii}. The relatively strong $\Delta z^P$ dependences of the various bispectra compared to that of GGG can be understood by comparing the effects of different $\Delta z^P$ on the weighting functions in Fig. \ref{fig:wabc}, which cause a strong decrease (or increase) in the amplitudes of $W^{GGI}$, $W^{GII}$, and $W^{III}$. The relative differences in the lines associated with the GGI (GGg) and GII (Ggg) bispectra are contributed to by both the different redshift dependencies of the two intrinsic alignment models, as well as the stronger dependence of the peak positions of $W^{GGI}$ and $W^{GII}$ on redshift separation in Fig. \ref{fig:wabc}, which decreases the accuracy of the approximation in Eqs. \ref{eq:ggirel} and \ref{eq:giirel}.}\label{fig:ratio}
\end{figure}

We can apply the $f_{GGG}$ calculated in this way from Method 1 to compare with the results of Method 3. We find that for $\ell=1000$ and $\bar{z^P}=1$, this results in $f_{GGG}=0.67$. Using this value, we find that the approximation in Eq. \ref{eq:taylor} is accurate to within 1\% at $\Delta z^P=0.3$. This is demonstrated in Fig. \ref{fig:ratio}, where the right side of Eq. \ref{eq:taylor} is plotted as the dotted line labeled GGG, which nearly overlaps the left side, plotted as the solid line labeled GGG. In general, we find $f_{GGG}\in(0.5,1.0)$ for $\ell\in(40,4000)$ at $\bar{z}^p=1$. Alternatively we can fit Eq. \ref{eq:taylor} to the calculated $B^{GGG}$ and determine a best-fit $f_{GGG}$. This results in a slightly lower value of $f_{GGG}$ with greatly improved accuracy, to within 0.1\%.

\subsubsection{Redshift separation dependence in $B^{III}$}\label{iii}

Unlike $W^{GGG}$, it is clear from Fig. \ref{fig:wabc} that $W^{III}$ is very dependent on $\Delta z^P$. However, the peak position in $z$ is determined by $\bar{z}^P$ and is insensitive to $\Delta z^P$, as long as the photo-z is sufficiently accurate. The peak's amplitude decreases with $\Delta z^P$, which explains the quick decrease in magnitude of $B^{III}$ seen in Fig. \ref{fig:ratio} as $\Delta z^P$ increases. This is a well-known result that is the basis for limiting measurements of the cosmic shear to only cross-correlations between thick photo-z bins in order to reduce the impact of the II, III, and GII signals.

This sharp peak in $W^{III}$ at $z_{\textrm{peak}}$, which is insensitive to $\Delta z^P$, allows us to make an approximation in the expression for $B^{III}$,
\begin{equation}
B^{III}(\ell;z^P_1,z^P_2,z^P_3)=\frac{B_{III}(k;\chi(z_{\textrm{peak}}))}{\chi^4(z_{\textrm{peak}})}\int_0^{\chi}W^{III}(\chi';\chi_1,\chi_2,\chi_3)d\chi',\label{eq:bsiii}
\end{equation}
where the dependence on $\Delta z^P$ is contained within the final integral. In the limit of perfect photo-z information, this $\Delta z^P$ dependence is exact. It is now clear that since $W^{III}=W^{ggg}$, we can relate $B^{III}$ and $B^{ggg}$ as
\begin{equation}
B^{III}(\ell;z^P_1,z^P_2,z^P_3)=A_{III}(\ell;z^P_1,z^P_2,z^P_3)B^{ggg}(\ell;z^P_1,z^P_2,z^P_3),\label{eq:gggiii}
\end{equation}
where $A_{III}\equiv B_{III}(k;\chi(z_{\textrm{peak}}))/B_{ggg}(k;\chi(z_{\textrm{peak}}))$. We propose to work with this relationship because it is in practice more accurate than Eq. \ref{eq:bsiii}, and because the same lensing survey will measure $B^{ggg}$ in addition to the shear bispectrum. This tells us directly the $\Delta z^P$ dependence in $B^{III}$ without any knowledge of the photo-z PDF or intrinsic alignment being necessary. In the same way, we can construct the following relationships as well:
\begin{eqnarray}
B^{IIg}(\ell;z^P_1,z^P_2,z^P_3)&=A_{IIg}(\ell;z^P_1,z^P_2,z^P_3)B^{ggg}(\ell;z^P_1,z^P_2,z^P_3),\label{eq:gggiii2}\\
B^{Igg}(\ell;z^P_1,z^P_2,z^P_3)&=A_{Igg}(\ell;z^P_1,z^P_2,z^P_3)B^{ggg}(\ell;z^P_1,z^P_2,z^P_3).\label{eq:gggiii3}
\end{eqnarray}
Figure \ref{fig:ratio} compares the $\Delta z^P$ dependence of $B^{III}$, $B^{IIg}$, $B^{Igg}$, and $B^{ggg}$ and shows that Eqs. \ref{eq:gggiii}-\ref{eq:gggiii3} are accurate to within 1\% at $\Delta z^P=0.2$. As with $f_{GGG}$, we will treat $A_{III}$, $A_{IIg}$, and $A_{Igg}$ as free parameters in order to avoid modeling uncertainty in the self-calibration discussed in Sec. \ref{cal}.

\subsubsection{Redshift separation dependence in $B^{GII}$}\label{gii}

We find that the sum $W^{GII}+W^{IGI}+W^{IIG}$ behaves very similarly to $W^{III}$, as shown in Fig. \ref{fig:wabc}, except for a slightly greater dependence in peak position due to $\Delta z^P$ and much less dependence on $\Delta z^P$ in peak amplitude. We thus propose a similar relationship to that described in Eq. \ref{eq:gggiii} for $B^{III}$, which is further motivated by the $\Delta z^P$ dependence of $B^{GII}+B^{IGI}+B^{IIG}$ in Fig. \ref{fig:ratio}, when compared to $B^{Ggg}+B^{gGg}+B^{ggG}$. Leaving out explicit dependence on $\ell$ and $z^P_1,z^P_2,z^P_3$, this relationship is approximately
\begin{equation}
B^{GII}+B^{IIG}+B^{IIG}\approx A_{GII}\left[B^{Ggg}+B^{gGg}+B^{ggG}\right].\label{eq:giirel}
\end{equation}
Due to the peak being less sharp as compared to that for $B^{III}$ and there being a stronger dependence on $\Delta z^P$ in the peak position, this relationship is not as exact as Eqs. \ref{eq:gggiii}-\ref{eq:gggiii3}. We still find a large degree of accuracy, however, with Eq. \ref{eq:giirel} being accurate to within 5\% at $\Delta z^P=0.2$.

\begin{figure}
\center
\includegraphics[angle=270,scale=0.65]{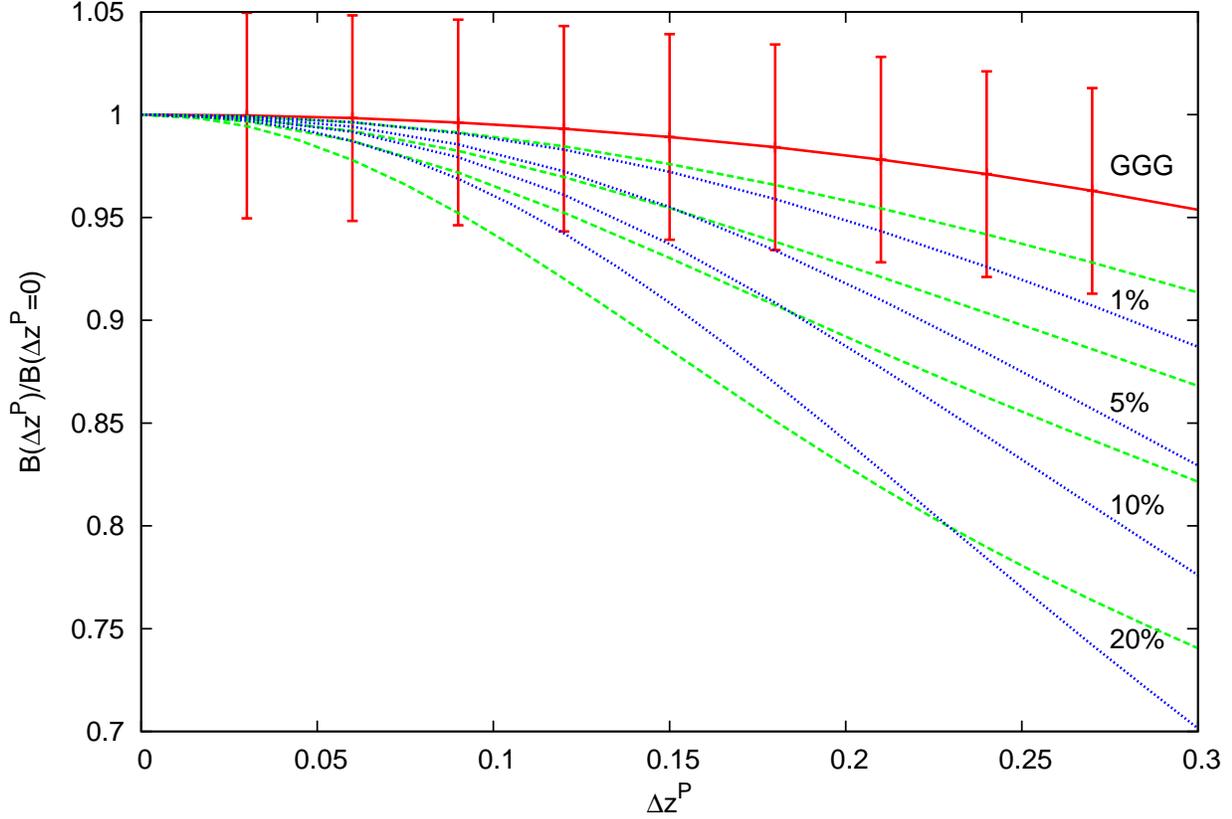}%
\caption{The ratio $B^{(1)}(\Delta z^P)/B^{(1)}(\Delta z^P=0)$, where $B^{(1)}=B^{GGG}+B^{GGI}+B^{GII}+B^{III}$, and the intrinsic alignment signal is modeled with both the SB09 model (dashed) and toy model with $a=1/2$ (dotted) for $\ell=1000$ and $\bar{z^P}=1.0$. From top to bottom, levels of total intrinsic alignment contamination $|B^{(1)}-B^{GGG}|/B^{(1)}$ of 1, 5, 10, and 20\% are shown. We are interested in the total absolute impact of the intrinsic alignment as a fraction of $B^{(1)}$, since this (and not the individual intrinsic alignment components) is what impacts cosmological information. For comparison we show $B^{GGG}$ (solid) with the expected minimum measurement uncertainty for an LSST-like survey. At this level of uncertainty, we find it is possible to identify the presence of even a small intrinsic alignment contamination of a few percent at $\Delta z^P=0.2$.}\label{fig:main}
\end{figure}

\subsubsection{Redshift separation dependence in $B^{GGI}$}\label{ggi}

Unlike $W^{III}$ and $W^{GII}+W^{IGI}+W^{IIG}$, the peak of the sum $W^{GGI}+W^{GIG}+W^{IGG}$ does clearly depend on $\Delta z^P$ as shown in Fig. \ref{fig:wabc}, though with a greater dependence in peak amplitude on $\Delta z^P$ compared to the sum $W^{GII}+W^{IGI}+W^{IIG}$ and $W^{III}$, instead increasing in amplitude with decreasing $\Delta z^P$. There is also a non-negligible contribution from the fraction $p_{\textrm{cat}}$ of outlier galaxies in Eq. \ref{eq:pdf} around $z=0.5$. However, from Fig. \ref{fig:ratio}, we still find a similar dependence on $\Delta z^P$ between $B^{GGI}+B^{GIG}+B^{IGG}$ and $B^{GGg}+B^{GgG}+B^{gGG}$, so we propose the relationship 
\begin{equation}
B^{GGI}+B^{GIG}+B^{IGG}\approx A_{GGI}\left[B^{GGg}+B^{GgG}+B^{gGG}\right].\label{eq:ggirel}
\end{equation}
While this relationship is much less accurate than Eqs. \ref{eq:gggiii}-\ref{eq:giirel}, due to the assumption that $z_{\textrm{peak}}$ is constant being less valid for GGI, it still is accurate to within 5\% at $\Delta z^P=0.1$ and 20\% at $\Delta z^P=0.2$. Thus if we wish to work at the accuracy possible for the 2-point case as shown by \cite{23}, where at $\Delta z^P=0.2$ the largest inaccuracy is about 10\% in the GI term, we must limit ourselves to $\Delta z^P\approx0.13$. Alternatively, we can use the limit proposed by Zhang of $\Delta z^P\approx0.2$, but with a loss of accuracy in Eq. \ref{eq:ggirel} of a factor of 2 when compared to the 2-point GI-Gg relationship.

We can see from Fig. \ref{fig:ratio} that the accuracy of Eq. \ref{eq:ggirel} could be improved by making assumptions about the redshift dependency of the intrinsic alignment or by measuring it independently. We could also include in the coefficient $A_{GGI}$ an extra dependency on redshift, instead of the simple scaling relationship we have assumed, which could account for redshift dependent sources of error like the changing position of $z_{\textrm{peak}}$ for GGI. This would introduce more free parameters into the self-calibration but would also allow the use of larger $\Delta z^P$. The choice of whether to include either intrinsic alignment model assumptions, additional measurements of redshift dependencies in the intrinsic alignment, additional free parameters in the self-calibration, or some combination of these or other assumptions will impact both the choice of upper $\Delta z^P$ and the ultimate performance of the self-calibration, but this choice must be made with the requirements of a particular lensing survey in mind. For this reason, we will limit our discussion to the basic framework developed in this section when addressing the performance of the self-calibration in Sec. \ref{perf}.

\section{Intrinsic alignment self-calibration}\label{cal}

We can now consider a means to self-calibrate the intrinsic alignment signal, by using the four measurable bispectra between galaxy ellipticity and galaxy density in a weak lensing survey, with negligible magnification bias, as discussed in \cite{troxel}:
\begin{eqnarray}
B^{(1)}&=&B^{GGG}+B^{GGI}+\textrm{2 perm.}+B^{GII}+\textrm{2 perm.}+B^{III},\label{eq:1}\\
B^{(2)}&=&B^{GGg}+\textrm{2 perm.}+B^{GIg}+\textrm{2 perm.}+3B^{IIg},\label{eq:3}\\
B^{(3)}&=&B^{Ggg}+\textrm{2 perm.}+3B^{Igg},\label{eq:2}\\
B^{(4)}&=&B^{ggg}.\label{eq:4}
\end{eqnarray}
Equation \ref{eq:1} is the measured galaxy ellipticity-ellipticity-ellipticity bispectrum, which measures both cosmic shear and correlated intrinsic alignment. In the case of no intrinsic alignment contamination, $B^{(1)}$ is simply the shear bispectrum, $B^{GGG}$, and should be effectively independent of $\Delta z^P$ within measurement error. However, we expect $B^{(1)}$ to have a measurable dependence on $\Delta z^P$ at minimum survey error with even just a few percent or more contamination by intrinsic alignment due to the very strong $\Delta z^P$ dependence of the intrinsic alignment bispectra. We explore this in Fig. \ref{fig:main}, where we plot both $B^{GGG}$ and $B^{(1)}$ for 1, 5, 10, and 20\% levels of intrinsic alignment contamination for both the SB09 and toy model. We are interested in the total absolute impact of the intrinsic alignment as a fraction of $B^{(1)}$, since this (and not the individual intrinsic alignment components) is what impacts cosmological information, and so we measure the contamination as $|B^{(1)}-B^{GGG}|/B^{(1)}$. Error bars representing the expected minimum measurement uncertainty in an LSST-like survey are shown on $B^{GGG}$. The measurement uncertainty in $B^{GGG}$ is extrapolated from the expected error found by \cite{23} for the power spectrum of LSST, modifying the derivation of \cite{troxel} for the lensing bispectrum in a single redshift bin.

Given simultaneous measurements of Eqs. \ref{eq:1}-\ref{eq:4} in small redshift bins of size $\sim 0.01$ at 6 or more $\Delta z^P$, along with the relations in Eqs. \ref{eq:taylor} and \ref{eq:gggiii}-\ref{eq:ggirel}, we can now simultaneously reconstruct $B^{GGG}$, $B^{GGI}$, $B^{GII}$, and $B^{III}$ through the observables given in Eqs \ref{eq:1}-\ref{eq:4} and with free parameters $f_{GGG}$, $A_{GGI}$, $A_{GII}$, $A_{III}$, $A_{Igg}$, and $A_{IIg}$. In practice, it is more useful to measure the total contamination $B^{GGI}+B^{GII}+B^{III}$, which can be determined to higher accuracy due to a partial degeneracy between the three intrinsic alignment bispectra. As we show in Fig. \ref{fig:ratio}, this is because all three cause $B^{(1)}$ to decrease with $\Delta z^P$. $B^{III}$ and $B^{GII}+\textrm{perm.}$ clearly decrease with $\Delta z^P$, and though $B^{GGI}+\textrm{perm.}$ increases with $\Delta z^P$, its magnitude is negative, and so it also causes $B^{(1)}$ to decrease with $\Delta z^P$. Though we have concentrated on a particular $\ell$ and $\bar{z}^P$, $\Delta z^P$ has a weak dependence on scale and mean redshift, so that we expect the self-calibration to be generally applicable to a wide range of values.

\subsection{Performance of the self-calibration}\label{perf}

From Fig. \ref{fig:wabc}, we see that to good approximation $B^{\alpha\beta\gamma}$ samples the same cosmic volume for different $\Delta z^P$, which means that different $\Delta z^P$ should share the same cosmic variance as found for the power spectrum \citep{23}. Relative differences due to cosmic variance are thus $<\sqrt{2\pi^2}(\ell^3\Delta\ell^3f_{\textrm{sky}})^{-1/2}=1.4\times 10^{-5}\%(10^3/\ell)^{3/2}(10^2/\Delta\ell)^{3/2}f_{\textrm{sky}}^{-1/2}$, and so are negligible compared to the $>5\%$ change in $B^{(1)}$ over $\Delta z^P=0.2$ for intrinsic alignment contaminations of $>1\%$. However, the shot noise due to random galaxy shapes is large compared to the cosmic variance at large $\ell$, for small redshift bins $\sim 0.01$ with a typical number of galaxies $2.5\times 10^7$ at $\bar{z}^P=1.0$ for LSST \citep{zhan}. It must be controlled in order to reach the necessary precision in order to identify the $\Delta z^P$ dependence at low levels of intrinsic alignment contamination. This is complicated by the fact that unlike cosmic variance, shot noise is uncorrelated at different $\Delta z^P$. \cite{23} discusses a means to reduce the shot noise by averaging over larger $\ell$ bins at varying $\bar{z}^P$, since the $\Delta z^P$ dependence is weak across $\ell$ and $\bar{z}^p$ for the power spectrum. We find this to be accurate for the bispectrum as well, and expect this method to be applicable in the 3-point self-calibration in order to reduce the shot noise to manageable levels in the small photo-z bins necessary for the self-calibration. We also anticipate the self-calibration to be applicable in the presence of other errors which have different $\Delta z^P$ dependencies.

In order for the self-calibration to reconstruct the intrinsic alignment signal in this way, with some associated error which is due primarily to the inaccuracy in Eqs. \ref{eq:ggirel} \& \ref{eq:giirel}, we must be able to measure 6 or more $\Delta z^P$ in $\sim 0.01$ photo-z bins which are distinct from the measurement error. If we consider the expected measurement error for LSST shown in Fig. \ref{fig:main}, we would be able to achieve the necessary measurements to self-calibrate the intrinsic alignment signal for intrinsic alignment contaminations of $\sim 10\%$ or more for maximum $\Delta z^P=0.2$. If we instead accept a greater inaccuracy in Eqs. \ref{eq:ggirel} \& \ref{eq:giirel} at higher $\Delta z^P$, it would then be possible to measure smaller intrinsic alignment contaminations. The precise choice between maximum $\Delta z^P$ and error in the final intrinsic alignment measurement through the self-calibration will then be entirely dependent upon the specific capabilities of the survey as well as the goals of the measurement. For example, if the resulting intrinsic alignment bispectrum is intended to constrain models of structure formation or intrinsic alignment models, then a lower maximum $\Delta z^P$ might be imposed in order to achieve better accuracy in the intrinsic alignment measurement.

\subsection{Other sources of uncertainty}\label{other}

The results of the previous section are dependent upon the accuracy of our assumptions in the quantitative calculations regarding the performance of the self-calibration. For the bispectrum, we have used an approximate fitting formula derived from perturbation theory by \cite{sc} in our performance estimations, and we have chosen a specific set of intrinsic alignment models. We also use a deterministic approach to modeling the galaxy bias, which is not perfectly accurate in real galaxy distributions. Failures in either these assumptions or the accuracy of the models for the intrinsic alignment bispectra would lead to additional uncertainty in the expected performance of the self-calibration, which may impact its applicability depending on the degree to which the assumptions or models fail. However, \cite{35} has shown that it is possible to suppress the galaxy stochasticity to the 1\% level in some cases, which would be safely negligible compared to other sources of error we have discussed above. We have also chosen two very different models of the intrinsic alignment, in order to minimize bias in the resulting performance evaluation.

\section{conclusion}\label{conc}

The strong $\Delta z^P$ dependency of the intrinsic alignment signal for large bin size $\ge 0.2$ has been used to motivate the preference of cross-spectra and bispectra between redshift bins in order to reduce the intrinsic alignment contamination. This has previously been used as a means to neglect the III and GII bispectra and II spectrum in techniques developed to remove the intrinsic alignment contamination from the lensing signal using information between redshift bins \citep{20a,20b,20c,21,22,troxel}. In this work, we instead use this $\Delta z^P$ dependency within a single redshift bin of size $\ge0.2$ to self-calibrate the measured galaxy ellipticity-ellipticity-ellipticity bispectrum, reconstructing simultaneously not only the cosmic shear bispectrum (GGG), but also the intrinsic alignment GGI, GII, and III bispectra.

We first explore several means of defining and measuring the $\Delta z^P$ dependence in the lensing and intrinsic alignment bispectra $B^{GGG}$, $B^{GGI}$, $B^{GII}$, and $B^{III}$, and show in Fig. \ref{fig:ratio} the resulting $\Delta z^P$ dependency of the best method as well as the accuracy of the scaling relations in Eqs. \ref{eq:gggiii}-\ref{eq:ggirel}. These relate the intrinsic alignment to the directly measurable galaxy density through a set of simple scaling parameters. This is true for both the toy model and the SB09 model of intrinsic alignment clustering, which have very different dependencies on both scale and redshift. We further show that unlike the intrinsic alignment bispectra, $B^{GGG}$ is effectively independent of $\Delta z^P$, which means a measured $\Delta z^P$ dependence in the measured ellipticity bispectrum is a clear indication of the presence of intrinsic alignment.

Using the measured cross-correlation galaxy ellipticity and galaxy density bispectra in a weak lensing survey, as well as the scaling relations in Eqs. \ref{eq:gggiii}-\ref{eq:ggirel}, we propose a simple self-calibration method to simultaneously reconstruct the GGG, GGI, GII, and III bispectra. The proposed self-calibration method relies only on the information already gathered by a weak lensing survey and makes no assumptions on the modeling of intrinsic alignment or photo-z PDF, instead depending on the relationship between the intrinsic alignment and galaxy density signals. We explore the feasibility and proposed structure of the self-calibration, including limitations on maximum $\Delta z^P$ due to the inaccuracy of Eqs. \ref{eq:ggirel} \& \ref{eq:giirel}, and discuss some means of addressing it. We find that $\Delta z^P\le0.2$ is sufficient for accuracy in Eq. \ref{eq:ggirel} better than 20\%. The precision of survey measurements of the lensing bispectrum also limits the applicability of the self-calibration, but we find that for the expected measurement error in LSST, we can reconstruct the intrinsic alignment signal for contaminations of 10\% at $\Delta z^P=0.2$, or for even smaller intrinsic alignment contaminations, but with less accuracy in the reconstructed intrinsic alignment signal. Thus this would allow the self-calibration technique to significantly reduce the contamination of the intrinsic alignment to the weak lensing spectrum and bispectrum. However, further work is still necessary to precisely evaluate the quantitative performance of the self-calibration method in a realistic survey and its anticipated reduction in the effects of the intrinsic alignment contamination on cosmological study.

The proposed self-calibration is complimentary to existing proposals for estimating the intrinsic alignment contamination to the bispectrum, which instead depend on information between redshift bins. It can be combined with these other methods which make use of redshift tomography to better constrain the intrinsic alignment contamination without over-using the information contained within the survey. By allowing the full reconstruction of the GII and III bispectra in addition to the GGG and GGI bispectra, while not relying on assumptions of an intrinsic alignment model, it also presents a means of indirectly measuring the intrinsic alignment signal, which will be applicable to constraining proposed intrinsic alignment models and our understanding of structure formation.

\section*{Acknowledgments}
We thank P. Zhang for suggesting this project, L. King for useful comments, and A. Peel for proof-reading this paper. MI acknowledges that this material is based upon work supported in part by National Science Foundation under grant AST-1109667 and NASA under grant NNX09AJ55G, and that part of the calculations for this work have been performed on the Cosmology Computer Cluster funded by the Hoblitzelle Foundation.

\label{lastpage}

\end{document}